\documentstyle[12pt]{article}
\begin{document}

%
\voffset -.70in \hoffset -.19in
\oddsidemargin 0in \evensidemargin 0in
\marginparwidth .75in \marginparsep 7pt \topmargin 0in
\headheight 12pt \headsep .25in
\footheight 18pt \footskip .35in
\textheight 9.5in \textwidth 6.1in
\columnsep 10pt \columnseprule 0pt
%
%
\def\tilde{\widetilde}
\def\bar{\overline}
\def\hat{\widehat}
\def\*{\star}
\def\({\left(}		\def\BL{\Bigr(}
\def\){\right)}		\def\BR{\Bigr)}
\def\[{\left[}		\def\BBL{\Bigr[}
\def\]{\right]}		\def\BBR{\Bigr]}
\def\lb{\[}
\def\rb{\]}
%
%
\def\frac#1#2{{#1 \over #2}}		\def\x{ \otimes }
\def\inv#1{{1 \over #1}}
\def\half{{1 \over 2}}
\def\d{\partial}
\def\der#1{{\partial \over \partial #1}}
\def\dd#1#2{{\partial #1 \over \partial #2}}
\def\vev#1{\langle #1 \rangle}
\def\ket#1{ | #1 \rangle}
\def\bra#1{ \langle #1 |}
\def\rvac{\hbox{$\vert 0\rangle$}}
\def\lvac{\hbox{$\langle 0 \vert $}}
\def\comm#1#2{ \BBL\ #1\ ,\ #2 \BBR }
\def\2pi{\hbox{$2\pi i$}}
\def\e#1{{\rm e}^{^{\textstyle #1}}}
\def\grad#1{\,\nabla\!_{{#1}}\,}
\def\dsl{\raise.15ex\hbox{/}\kern-.57em\partial}
\def\Dsl{\,\raise.15ex\hbox{/}\mkern-.13.5mu D}
%
%
\def\th{\theta}		\def\Th{\Theta}
\def\ga{\gamma}		\def\Ga{\Gamma}
\def\be{\beta}
\def\al{\alpha}
\def\ep{\epsilon}
\def\la{\lambda}	\def\La{\Lambda}
\def\de{\delta}		\def\De{\Delta}
\def\om{\omega}		\def\Om{\Omega}
\def\sig{\sigma}	\def\Sig{\Sigma}
\def\vphi{\varphi}
%
%
\def\CA{{\cal A}}	\def\CB{{\cal B}}	\def\CC{{\cal C}}
\def\CD{{\cal D}}	\def\CE{{\cal E}}	\def\CF{{\cal F}}
\def\CG{{\cal G}}	\def\CH{{\cal H}}	\def\CI{{\cal J}}
\def\CJ{{\cal J}}	\def\CK{{\cal K}}	\def\CL{{\cal L}}
\def\CM{{\cal M}}	\def\CN{{\cal N}}	\def\CO{{\cal O}}
\def\CP{{\cal P}}	\def\CQ{{\cal Q}}	\def\CR{{\cal R}}
\def\CS{{\cal S}}	\def\CT{{\cal T}}	\def\CU{{\cal U}}
\def\CV{{\cal V}}	\def\CW{{\cal W}}	\def\CX{{\cal X}}
\def\CY{{\cal Y}}	\def\CZ{{\cal Z}}
%
%
\font\numbers=cmss12
\font\upright=cmu10 scaled\magstep1
\def\stroke{\vrule height8pt width0.4pt depth-0.1pt}
\def\topfleck{\vrule height8pt width0.5pt depth-5.9pt}
\def\botfleck{\vrule height2pt width0.5pt depth0.1pt}
\def\Zmath{\vcenter{\hbox{\numbers\rlap{\rlap{Z}\kern 0.8pt\topfleck}\kern
2.2pt
                   \rlap Z\kern 6pt\botfleck\kern 1pt}}}
\def\Qmath{\vcenter{\hbox{\upright\rlap{\rlap{Q}\kern
                   3.8pt\stroke}\phantom{Q}}}}
\def\Nmath{\vcenter{\hbox{\upright\rlap{I}\kern 1.7pt N}}}
\def\Cmath{\vcenter{\hbox{\upright\rlap{\rlap{C}\kern
                   3.8pt\stroke}\phantom{C}}}}
\def\Rmath{\vcenter{\hbox{\upright\rlap{I}\kern 1.7pt R}}}
\def\Z{\ifmmode\Zmath\else$\Zmath$\fi}
\def\Q{\ifmmode\Qmath\else$\Qmath$\fi}
\def\N{\ifmmode\Nmath\else$\Nmath$\fi}
\def\C{\ifmmode\Cmath\else$\Cmath$\fi}
\def\R{\ifmmode\Rmath\else$\Rmath$\fi}

\def\x{\stackrel{\otimes}{,}}
\def\Gr{ G_R }
\def\gr{ \CG_R }
\def\oti{ \otimes }
\def\otp{ {\oti_,} }
\def\debut{ \begin{eqnarray} }
\def\fin{ \end{eqnarray} }
\def\non { \nonumber\\ }
\def\bb{ $\bullet$}


\rightline{SPhT-92-134}
\vskip 2cm
\centerline{\LARGE An Introduction to Yangian Symmetries
\footnote{Talk given at the "Integrable Quantum Field Theories" conference
held at Come, Italy , September 13-19, 1992}.}
\vskip1cm
 \centerline{\large  Denis Bernard }
 \centerline{Service de Physique Th\'eorique de Saclay
\footnote[2]{\it Laboratoire de la Direction des Sciences de la
Mati\`ere du Commisariat \`a l'Energie Atomique.}}
\centerline{F-91191, Gif-sur-Yvette, France.}
\vskip 1cm
\font\petit=cmr10
\noindent
{\petit  We review some aspects of the quantum Yangians as symmetry algebras of
two-dimensional quantum field theories. The plan of these notes is the
following:\\
1 - The  classical Heisenberg model:\\
\bb Non-Abelian symmetries;
\bb The generators of the symmetries and the semi-classical Yangians;
\bb An alternative presentation of the
semi-classical Yangians;
\bb Digression on Poisson-Lie groups.\\
2 - The quantum Heisenberg chain:\\
\bb Non-Abelian symmetries and the quantum Yangians;
\bb The transfer matrix and an alternative presentation of the Yangians;
\bb Digression on the double Yangians.\\ }

\bigskip

\noindent Quantum integrable models are characterized by the existence
of commuting conserved charges which one is free to choose as Hamiltonians.
They usually also possess extra symmetries which are at the
origin of degeneracies in the spectrum. These symmetries are
often not a group but a quantum group. The Yangians \cite{Dr86} are quantum
group symmetries which (almost systematically) show up in integrable models
invariant under an internal non-Abelian group. The most standard example
of this kind is the Heisenberg model which we use in these notes for
presenting an introduction to the Yangian algebras. Other examples of
two-dimensional relativistic quantum field theories which are Yangian
invariant are provided  e.g. by sigma models, two dimensional massive
current algebras, principal chiral models with or without topological terms
\cite{LuPo78,B91}, etc...
Usually the infinite dimensional quantum symmetries are not
compatible with periodic boundary conditions. However more recently,
it has been realized that a variant of the Heisenberg chain with a long-range
interaction was Yangian invariant in a way compatible with
periodic boundary conditions, making eigenstates countable and the
degeneracy easy to describe \cite{Ha92}. It is tempting to
speculate that quantum integrable theories could be solved only using
their symmetry algebras in a way similar to the solutions of two-dimensional
conformal field theories. An introduction to quantum groups in relation
with lattice models can be found in ref. \cite{Pa92}. These notes arise
essentially from a joint work with Olivier Babelon, ref. \cite{BaBe92}.

\section{The classical Heisenberg model}

We start with the definition of the classical Heisenberg  model.
As a classical Hamiltonian system, we need to introduce the phase space
and its symplectic structure as well as the Hamiltonian. The classical
variables are the following spin variables $S(x)$:
\debut
S(x)=\sum_{i=1}^3 S^i(x) \sigma_i\quad
{\rm with}\quad \sum_{i=1}^3 S^i(x)^2=s^2 \nonumber
\fin
where the $\sigma_i$ are the Pauli matrices with $[\sig_i,\sig_j]=
2i\ep_{ijk}\sig_k$ and $tr(\sig_j\sig_k)=2\de_{jk}$.
Here and below, $s$ is a fixed real number.
The Poisson bracket are defined by
\debut
\{ S^i(x),S^j(y)\} = \epsilon^{ijk} S^k(x)\delta(x-y) \label{Ki}
\fin
They are symplectic thanks to the constraint $\sum_i S^i(x)^2 = s^2$.
The Hamiltonian is
\debut
H_1=-{1\over 4}\int_0^L dx~ tr(\partial_x S \partial_x S) \label{Kii}
\fin
The equations of motion deduced from this Hamiltonian are:
\debut
\d_t S
=-{\textstyle{i\over 2}} \BBL S ,\partial_x^2 S \BBR={\textstyle{i\over 2}}
\partial_x \BBL\ S_x ,S\ \BBR  \label{EHv}
\fin
Notice that these equations are the conservation laws for a $su(2)$-valued
current.
As is well known, the Heisenberg model is a completely integrable model.
Its integrability relies on the fact that the equations (\ref{EHv})
can be written as a zero curvature condition for an auxiliary linear
problem. Namely, consider the Lax connexion,
\debut
A_x& =& {i\over \lambda}S(x) \non
A_t& =&- {{2is^2}\over \lambda^2} S(x) +
{1\over{2\lambda}}[S(x),\partial_x S(x)]\label{EHvii}
\fin
then, the zero curvature condition, $\[ \d_t + A_t , \d_x + A_x \] = 0$,
is equivalent to the equations of motion.
Note that the Lax connexion (\ref{EHvii}) is an element of the $su(2)$ loop
algebra ${\tilde {su(2)}}=su(2)\otimes C\[\la,\la^{-1}\]$.

An important ingredient is the transfer matrix $T(x,\lambda)$;
it is defined as
\debut
T(x,\lambda)&=&P \exp\left[-\int_0^x A_x(y,\lambda) dy\right]
\label{EHviii}
\fin
The monodromy matrix $T(\lambda)$ is simply $T(L,\lambda)$.
 From its definition, $T(x,\lambda)$ is analytic in $\lambda$
with an essential singularity at $\lambda =0$. From eq.(\ref{EHviii}), we can
easily find an expansion around $\lambda =\infty $:
\debut
T(x,\lambda )= 1 -{\textstyle{i\over \lambda}}\int_0^x dy S(y)
 -{\textstyle{1\over \lambda^2}}\int_0^x dy S(y) \int_0^y dz S(z) + \cdots
\nonumber
\fin
This development in $1/\la$ has an infinite radius of convergence.
We will study it later in more details in relation with the
non-Abelian symmetries of the model.

As is well known, the importance of the monodromy matrix lies in
the fact that one can calculate the Poisson bracket of its  matrix
elements. One finds \cite{Sk79}:
\debut
\Bigl\{ T(\lambda) \x T(\mu)\Bigr\}={\textstyle{1\over 2}}
\BBL\ r(\lambda,\mu)\ ,\ T(\lambda)\otimes T(\mu)\ \BBR \label{EHix}
\fin
where
\debut
r(\lambda ,\mu)= \inv{ \lambda-\mu}\ {\sum_i \sigma_i \otimes \sigma_i}
\label{EHx}
\fin
 From this result, it follows that $tr(T(\lambda))$ is a generating function
for quantities in involution.

Assuming periodic boundary conditions, the monodromy matrix $T(\lambda)$
can be written as follows:
$T(\lambda)= \cos P_0(\lambda)~{\rm Id} +i \sin P_0(\lambda)~ M(\lambda)$,
with $M(\lambda)$ traceless.
Therefore, the trace of the transfer matrix is:
$tr(T(\lambda ))=2 \cos P_0(\lambda)$, and
we can use $P_0(\lambda)$ as a generating function
for the commuting conserved quantities. The local conserved charges
are found by expanding not around $\lambda=\infty$ but around
$\lambda= 0$:
\debut
P_o(\lambda) = -\frac{sL}{\lambda}
+ \sum_{n=0}^{\infty} \lambda^n\ I_n \nonumber
\fin
The quantities $I_n$ are integral of local densities.
The first two, $I_0$ and $I_1$, correspond to momentum and energy respectively.

\subsection{ Non-Abelian symmetries.}

Clearly, since the Hamiltonian is a $su(2)$ scalar, the equations
of motion are $su(2)$ invariant. Therefore, for any element $v\in su(2)$,
the transformation,
\debut
\delta_v\ S(x)\ =\ i\BBL\ v\ ,\ S(x)\ \BBR \nonumber
\fin
is a symmetry of the equations of motion.
Actually, the symmetry group is much bigger.
For any $v\in\ su(2)$ and any non-negative integer $n$,
the transformations $S(x)\to\ \delta_v^nS(x)$  defined by:
\begin{eqnarray}
\delta^0_v\ S(x)\ &=&\ i\BBL\ v\ ,\ S(x)\ \BBR \non
\delta^n_v\ S(x)\ &=&\ i\BBL\ Z^n_v(x)\ ,\ S(x)\ \BBR\label{dt4}
\end{eqnarray}
where the functions $Z^k_v(x)$ are  recursively computed  by
\begin{eqnarray}
\partial_x Z^k_v(x) +i \BBL S(x),Z^{k-1}_v(x)\BBR =0\qquad;
\qquad Z_v^0 =v \label{dt3}
\end{eqnarray}
are symmetries of the equations of motion.

Moreover, these transformations form a representation of
the loop algebra (more precisely of the sub-algebra
$su(2)\otimes C[\lambda]$ of the $su(2)$ loop algebra); i.e. we have:
\debut
\BBL\ \delta_v^n\ , \delta_w^m\ \BBR = \ \delta_{[v,w]}^{n+m}
\fin
for any non-negative intergers $n$ and $m$, and any $v,\ w \in su(2)$.
In other words, the symmetry group is the loop group
(more precisely the sub-group of the loop group which consists
of loops regular at zero).

This can be proved as follows.
For any $v\in su(2)$, let us define the functions $Z^k_v(x)$ by:
\begin{eqnarray}
(TvT^{-1})(x,\la) =\sum_{k=0}^\infty \lambda^{-k} Z^k_v(x) \nonumber
\end{eqnarray}
where we have used the $(\frac{1}{\lambda})$ expansion of the transfer matrix.
The differential equations satisfied by these functions are consequences
of those satisfied by the transfer matrix $T(x,\la)$. For any positive
integer $n$, we set:
$\Theta_v^n(x,\lambda) =i\sum_{k=0}^n \lambda^{n-k} Z_v^k(x)$.
Consider now the following gauge transformations acting on the Lax connexion:
\debut
\delta_v^n A_x &=&- [A_x, \Theta_v^n] - \partial_x \Theta_v^n \nonumber \\
\delta_v^n A_t &=&- [A_t, \Theta_v^n] - \partial_t \Theta_v^n \nonumber
\fin
By construction, these transformations preserve the zero curvature
condition, since they are gauge transformations.
Therefore, they will be symmetries if
the form of the components, $A_t$ and $A_x$, of the  Lax connexion is
preserved.
It is a simple exercise to check this fact; e.g. for $A_x$  we have:
\begin{eqnarray}
\delta_v^n A_x &=& \lambda^{-1} [ S, Z^n_v ] -
 i\sum_{k=0}^{n-1} \left\{ \partial_x Z_v^{k+1}
 +i [S, Z_v^k] \right\}\lambda^{n-k-1} \nonumber \\
 &=& \lambda^{-1}  [ S, Z^n_v ] \nonumber
 \end{eqnarray}
The last sum vanishes by vitue of eq.(\ref{dt3}) and we are left with
$\delta^n_v S(x) = i\[ Z^n_v(x),S(x)\]$.
One can check similarly that the form of $A_t$ is unchanged, and
its variation is compatible with eq.(\ref{dt4}).
This proves that eqs.(\ref{dt4}) define symmetries of the equations
of motion.

To these symmetries correspond an infinite number of conserved currents.
Indeed, as we already remarked the equations of motion
have the form of a conservation law $\partial_t J_t - \partial_x J_x =0$
with $ J_t =S$ and $J_x={i\over 2}[S_x,S]$.
Since the transformations are symmetries,
transforming this local current produces new
currents which form an infinite multiplet of currents:
\begin{eqnarray}
J_t^{n,v} &=& \delta^n_v S = i[Z_v^n,S ] \nonumber \\
J_x^{n,v} &=&- {\textstyle{1\over 2}}[ \partial_x [Z_v^n,S ],S]
-{\textstyle{1\over 2}} [S_x ,[Z_v^n,S]] \label{Ecur}
\end{eqnarray}
for any $n\geq 0$ and $v\in su(2)$.
Note that these currents are non-local. By construction they are conserved:
$ \partial_t J^{n,v}_t - \partial_x J^{n,v}_x =0$.
To them correspond charges which are defined by:
\begin{eqnarray}
Q^n_v =\int_0^L J_t^{n,v} (x) dx  = Z_v^{n+1}(L) \label{Echar}
\end{eqnarray}
Since the currents are non local, the charges are not conserved. We have
\begin{eqnarray}
{d\over dt}Q_v^n = J_x^{n,v} (L) - J_x^{n,v} (0) \nonumber
\end{eqnarray}
In the infinite volume limit ($L\to\infty$), with an appropriate choice
of the boundary conditions the charges can eventually be conserved.
Even if they are not conserved, they are nevertheless important because,
as we will see, they are the generators of the non-Abelian  transformations.

\subsection{ The generators of the symmetries and the
semi-classical Yangians.}

We now discuss in which sense the non-local charges $Q_v^n$
are the generators of the non-Abelian symmetries (\ref{dt4}); i.e.
in which sense the infinitesimal transformation laws (\ref{dt4}) of
the spin variables are given by Poisson brackets between
the non-local charges and the dynamical variables $S(x)$.
As we will see, contrary to what happens for symplectic
transformations, the infinitesimal variations $\delta_v^nS(x)$
are not linearly generated by the charges.

The charges $Q_v^n$ take values in the Lie algebra $su(2)$.
Let us introduce their components, $Q_{ij}^n$ and
$Q_i^n$, in the basis of the Pauli matrices $\sigma_i$:
\debut
Q^n_{ij} =\half\ tr\({ Q_{\sigma_i}^n\ \sigma_j}\)\quad
{\rm and}\quad
Q^n_i = \sum_{j,k=0}^3 \ep_{ijk} Q_{jk}^n \label{CIv}
\fin
In particular,  for $n=0$ and $n=1$, we have the simple formula:
\debut
Q_i^0 &=& 4 \int_0^L\ dx\ S^i(x) \non
Q_i^1 &=& 4 \int_0^L\ dx \int_0^x dy\ \ep^{ijk}\  S^j(x) S^k(y) \label{CIvi}
\fin
Notice that $Q_i^0$ are local whereas $Q_i^1$ are not.
They generate the transformations $\delta_i^0S(y)$ and
$\delta^1_iS(y)$ in the following way:
\begin{eqnarray}
\delta^0_{i} S(y)&=&{\textstyle{1\over 2}}  \{ Q^0_i,S(y) \} \non
 \delta_i^1 S(y)&=&{\textstyle{1\over 2}} \{Q_i^1 ,S(y) \}
-{\textstyle{1\over 8}} \epsilon^{ijk} Q_j^0 \{ Q_k^0 ,S(y) \}
\label{EZv}
\end{eqnarray}
The first equation in (\ref{EZv}) implies that the $su(2)$
symmetry $\delta^0_iS(y)$ is a symplectic action as it should
be. The non-linearity in the second equation is the sign that these
transformations are not symplectic but is characteristic to
Lie-Poisson actions, a notion which generalize the notion of symplectic
actions.

The non-linearity in eq. (\ref{EZv}) has another echo. The Poisson algebra
of the charges $Q_i^0$ and $Q^1_i$ is not the $su(2)$ loop algebra
but a deformation of it. Indeed recall that the $su(2)$ loop algebra
can be presented as the associative algebra generated by the
elements $\delta^0_i$ and $\delta^1_i$ satisfying the following relations:
\debut
\BBL \delta_i^0,\delta_j^0\BBR &=&  \ep_{ijk} \delta_k^0 \non
\BBL \delta_i^0,\delta_j^1\BBR &=&  \ep_{ijk} \delta_k^1 \label{CIx}\\
\[\delta^1_i,\[\delta^1_j,\delta^0_k\]\]
&-&\[\delta^0_i,\[\delta^1_j,\delta^1_k\]\] =0 \non
\[\[\delta^1_i,\delta^1_j\],\[\delta^0_k,\delta^1_l\] \]
&+&\[\[\delta^1_k,\delta^1_l\],\[\delta^0_i,\delta^1_j\]\]=0 \nonumber
\fin

On the other hand, given the explicit expressions of the charges
$Q_i^0$ and $Q_i^1$ in terms of the spin variables, one can
compute their Poisson brackets and check that they satisfy the following
relations:
\debut
\{ Q_i^0,Q_j^0\} &=& 4 \ep_{ijk} Q^0_k \non
\{ Q_i^0,Q_j^1\} &=& 4 \ep_{ijk} Q^1_k \label{CIxi}\\
\{Q^1_i,\{Q^1_j,Q^0_k\}\} -\{Q^0_i,\{Q^1_j,Q^1_k\}\} &=&
A_{ijk}^{lmn} Q^0_l Q^0_m Q^0_n \non
\{\{Q^1_i,Q^1_j\},\{Q^0_k,Q^1_l\} \}
+\{\{Q^1_k,Q^1_l\},\{Q^0_i,Q^1_j\}\} &=&
8( A_{ija}^{mnp}\ep_{kla} + A^{mnp}_{kla} \ep_{ija}) Q^0_m Q^0_n Q^1_p
\nonumber
\fin
with $A_{ijk}^{lmn}= {\textstyle{2\over
3}}\ep_{ila}\ep_{jmb}\ep_{knc}\ep^{abc}$

One sees that these relations form a deformation of those defining
the $su(2)$ loop algebra. Therefore the Poisson algebra of the charges
is a deformation of the $su(2)$ loop algebra.
It is called the semi-classical $su(2)$ Yangian.
There is no extra relation between the charges $Q^0_n$ and $Q^1_k$ since
there is no extra relation between the generators $\de^0_n$ and $\de^1_k$
in the $su(2)$ loop algebra.
\par

All the non-local charges $Q_{ij}^n$ can be expressed as multiple
Poisson brackets between the two first charges $Q_i^0$ and $Q_i^1$.
Therefore, the Poisson algebra of the symmetries is generated by these two
charges.
(This fact is more easily proved by introducing the
transfer matrix and its Poisson brackets as it will be done
in the next section.)

Also, since the $su(2)$ loop algebra is generated only by
$\delta^0_i$ and $\delta^1_i$, all the infinitesimal transformations
$\delta^n_iS(y)$ can be expressed non-linearly in terms of
the non-local charges $Q_i^0$ and $Q_i^1$, or alternatively,
 in terms of the charges $Q^n_{ij}$. A close expression is:
\begin{eqnarray}
\delta^n_{i} S(y)= {\textstyle{1\over 2}} \{ Q^n_i , S(y) \}
 +{\textstyle{1\over 2}}\sum_{p=0}^{n-1}\left[ Q^{n-p-1}-tr(Q^{n-p-1})Id
\right]_{ik} \delta_k^p S(y)
\nonumber
\end{eqnarray}
This equation can be interpreted in two different ways: (i)
this is a system of equations allowing to express recursively $\delta_v^n
S(y)$ in terms of Poisson brackets; or (ii) it allows to express the
transformed spins $\de_v^nS$ in a non-linear way in terms of the charges $Q_i$.

\subsection{The transfer matrix and an alternative presentation
of the semi-classical Yangians.}

We now show that the knowledge of the charges is equivalent to the
data of the monodromy matrix $T$. Let us introduce a notation for
the matrix elements of the monodromy matrix:
\begin{eqnarray}
T(\lambda)=\pmatrix{A(\lambda )& B(\lambda ) \cr C(\lambda ) & D( \lambda ) }
\qquad;\qquad Det\ T(\lambda) = AD-BC=1  \nonumber
\end{eqnarray}
Recall that from eq. (\ref{CIv}), we have:
\debut
Q_{ij}(\lambda) &=& \delta_{ij}+ \sum_{n=0}^\infty \lambda^{-n-1} Q^n_{ij} =
\half tr( T \sigma_i T^{-1} \sigma_j ) \non
Q_i(\lambda) &=& \sum_{n=0}^\infty \lambda^{-n-1} Q^n_i =
\ep_{ijk} Q_{jk}(\lambda) \nonumber
\fin
Therefore, the quantities $Q_{ij}(\lambda)$ and $Q_i(\lambda)$ are quadratic
functions of the matrix elements of $T(\lambda)$.
We also point out that it is possible to express the charges $Q_{ij}^n$
in terms of $Q_i^n$.

But, one can invert these relations and express $T(\lambda)$ in terms
of the generating functions $Q_i(\lambda)$.
The relation between the transfer matrix and the
non-local charges $Q_i(\la)$ is:
\begin{eqnarray}
T(\lambda)=
{\inv{2}}W(\lambda)\ Id -{\frac{i}{2}} W^{-1}(\lambda)
\sum_i Q_i(\lambda) \sigma_i
\nonumber
\end{eqnarray}
with $ W(\lambda)= \sqrt{ 2 + \sqrt{4-\vec{Q}^2(\lambda)} }$.
\par
So, the charges $Q_i(\lambda)$ contains the same amount of
information as the monodromy matrix.
It is interesting in this context to examine more closely the relation
between the $Q_i(\lambda)$ and the matrix elements of $T(\la)$.
We have
\begin{eqnarray}
Q_+(\lambda) &=& Q_1(\lambda) +i Q_2(\lambda) = 2i
W(\lambda)\  C(\lambda) \label{nlc1} \\
Q_-(\lambda) &=& Q_1(\lambda) -i Q_2(\lambda) = 2i
W(\lambda)\  B(\lambda) \label{nlc2}
\end{eqnarray}
Moreover, the trace of the transfer matrix is:
$tr~T(\lambda)=W[\vec{Q}^2(\lambda)]$.
Therefore, $\vec{Q}^2(\lambda)$ is also a generating
function for commuting quantities. Its relation with the generating
function $P_0(\la)$ is:
\debut
\vec{Q}^2(\la)\ =\ 4\sin^2\BL 2P_0(\la)\BR \nonumber
\fin
However, expanding $\vec{Q}^2(\la)$ around $\la=\infty$ gives
non-local commuting quantities while expanding $P_0(\la)$ around
$\la=0$ gives the local commuting quantities. The links between
them are hidden in the subtilities of the analytic properties
of the monodromy matrix (these analytic properties should be encoded
into the representation theory of the Yangian).
\par

Since, the transfer matrix encodes the same amount of information
as the charges, the semi-classical $su(2)$ Yangian (\ref{CIxi}) can
also be presented in terms of the transfer matrix:
the semi-classical $su(2)$ Yangians is the Poisson algebra generated by
the transfer matrix, $T(\lambda)$, of unit determinant
and with Poisson brackets:
\debut
\Bigl\{ T(\lambda) \x T(\mu)\Bigr\}={\textstyle{1\over 2}}
\BBL\ r(\lambda,\mu)\ ,\ T(\lambda)\otimes T(\mu)\ \BBR \label{EHixb}
\fin
The $\frac{1}{\lambda}$ expansion of $T(\lambda)$ is implicitely
assumed in the definition. For the components $t^{(n)}_{ab}$ of $T(\lambda)$,
defined by $T(\lambda)_{ab}=\delta_{ab}+\sum_{n=0}^{\infty}
t^{(n)}_{ab}\lambda^{-n-1}$, we get:
\debut
\Bigl\{  t^{(n)}_{ab},t^{(m)}_{cd}  \Bigr\}
= \delta_{cb}t^{(n+m)}_{ad}- \delta_{ad} t^{(n+m)}_{cb}
+\sum_{p=0}^{n-1}\BL  t^{(m+p)}_{ad}t^{(n-1-p)}_{cb}
- t^{(n-1-p)}_{ad}t^{(m+p)}_{cb} \BR
\label{CIxv}
\fin
In particular, this last relation shows that the Poisson algebra is
effectively generated by the two first charges $Q_i^0$ and $Q_i^1$.

Finally, let us describe how the monodromy matrix generates
the transformations (\ref{dt4}).
By an explicit computation of the Poisson brackets between
the monodromy matrix and the spin variables, one can easily check
that the variation $\delta_v^n S(y)$ of the spin variables are
given by following formula:
\begin{eqnarray}
\delta_v^n S(y)=\ i\oint {{d\lambda }\over{2i\pi}} \lambda^n\
tr_1\BL\ (vT^{-1}(\lambda)\otimes 1)\Bigl\{
T(\lambda)\otimes 1 , 1\otimes S(y) \Bigr\}\ \BR \label{CIxvi}
\end{eqnarray}
Here $tr_1$ denotes the trace over the first space in the tensor product.
For $n=0$ or $1$, eq. (\ref{CIxvi}) is equivalent to eq. (\ref{EZv}).
It indicates that $T(\lambda) $ is the generator of
the non-Abelian symmteries and characterizes the transformations
$S(y)\to \delta^n_vS(y)$ as Lie-Poisson actions.

\subsection{Digression on Poisson-Lie groups.}

In this section we present a few basic facts about Lie-Poisson
actions \cite{Dr86,Se83}. Let $M$ be a sympletic manifold.
We denote by $\{\ ,\ \}_M$ the Poisson bracket in $M$.

Before describing Lie-Poisson actions,
we recall some well known facts about Hamiltonian actions.
Let $H$ be a Lie group and $\CH$ its Lie algebra. The action of
a one parameter subgroup $(h^t)$ of $H$ is said to be symplectic if for any
functions $f_1$ and $f_2$ on $M$ ,
\begin{eqnarray}
\{ f_1(h^t.x), f_2(h^t.x)\}_M\ =\ \{f_1,f_2\}_M(h^t.x) \label{EIi}
\end{eqnarray}
Introducing the vector field $X$ on $M$ corresponding to the infinitesimal
action, $\de_X.f(x) = \frac{d}{dt}f(h^t.x)|_{t=0}$, the condition (\ref{EIi})
becomes:
\begin{eqnarray}
\{ \de_X.f_1 , f_2 \}_M\ +\ \{ f_1 , \de_X.f_2 \}_M\ =\
\de_X.\{f_1,f_2\}_M \nonumber
\end{eqnarray}
We have the standard property that the action of any one parameter subgroup
of $H$ is locally Hamiltonian. This means that there
exists a function $H_X$, locally defined on $M$, such that:
\begin{eqnarray}
\de_X.f\ =\ \{H_X, f\}_M \label{EIiii}
\end{eqnarray}
The proof is standard.
The global existence of $H_X$ is another state of affair.
The Hamiltonians $H_X$ are used to define the moment map.
These properties generalize to Lie-Poisson actions.

A Poisson- Lie group $H$ is a Lie group equipped with a Poisson
structure such that the multiplication in $H$ viewed as map
$H\times H \to H$ is a Poisson mapping.
Let us be more explicit. Any Poisson bracket $\{,\}_H$ on a Lie
group $H$ is uniquely characterized by the data of a
$\CH\oti\CH$-valued function: $h\in H\to \eta(h)\in\CH\oti\CH$.
Indeed, introducing a basis $(e_a)$ of $\CH$, the Poisson
bracket  for any functions
$f_1$ and $f_2$ on $H$ can be written as :
\begin{eqnarray}
\{f_1,f_2\}_H(h)=\sum_{a,b}\ \eta^{ab}(h)(\nabla^R_af_1)(h)
		(\nabla^R_bf_2)(h) \label{Epi}
\end{eqnarray}
where $\eta(h)=\sum_{a,b}\eta^{ab}(h)e_a\oti e_b$ and
$\nabla^R_a$ is the right-invariant vector field corresponding
to the element $e_a\in\CH$:
$\nabla^R_a f(h)= { {d\over{dt}}f(e^{te_a}h)}\Big\vert_{t=0}$.
The antisymmetry of the Poisson bracket (\ref{Epi})
requires $\eta_{12}=-\eta_{21}$, and the Jacobi identity is equivalent
to a quadratic relation for $\eta$ which can be easily written down.
The Lie Poisson property of the Poisson brackets (\ref{Epi}) is the
requirement that they transform covariantly under the multiplication
in $H$; It requires that $\eta(h)$ is a cocycle \cite{Dr86}:
$\eta(hg) = \eta(h) + Ad\, h\cdot \eta(g)$.

The bracket $\{,\}_H$ can be used to define a Lie algebra
structure on $\CH^*$ by
$\[ d_ef_1, d_ef_2 \]_{\CH^*}$ $= d_e\{f_1,f_2\}_H$,
with $d_ef\ \in\CH^*$ the differential of the function $f$
on $H$ evaluated at the identity of $H$.
In a basis $(e^a)$ in $\CH^*$, dual to
the basis $(e_a)$ in $\CH$, the differential at the identity can
written as $d_ef=\sum_a e^a(\nabla^L_af)(e)\in\CH^*$ where
$\nabla^L_a$ are the left-invariant vector fields on $H$, and
the Lie structure in $\CH^*$ is:
\debut
\[e^a,e^b\]_{\CH^*}\ =\ f^{ab}_c\ e^c \label{Epv}
\fin
where the structure constants are $f^{ab}_c=(\nabla^L_c\eta^{ab})(e)$.
The Lie bracket eq.(\ref{Epv}) satisfies the Jacobi identity thanks to the
Jacobi identity for the Poisson bracket in $H$.
We denote by $H^*$ the Lie group with Lie algebra $\CH^*$.

The action of a Poisson-Lie group on a symplectic manilfold is a Lie-Poisson
action if the Poisson brackets transform covariantly;
i.e. if for any $h\in H$ and any function $f_1$ and $f_2$ on $M$,
\begin{eqnarray}
\{f_1(h.x),f_2(h.x)\}_{H\times M}\ =\
\{f_1,f_2\}_M(h.x) \label{EIviii}
\end{eqnarray}
The Poisson structure on $H\times M$ is the product Poisson structure.

Let $X\in\CH$ and denote also by $\de_X$ the vector field on $M$ corresponding
to the infinitesimal transformation generated by $X$.
Introducing two dual basis of the Lie algebras $\CH$ and $\CH^*$,
$e_a\in\CH$ and $e^a\in\CH^*$ with $<e^a,e_b>=\delta^a_b$,
where $<,>$ denote the pairing between $\CH$ and $\CH^*$,
eq. (\ref{EIviii}) becomes:
\begin{eqnarray}
\{\de_{e_a}.f_1,f_2\}_M\ +\
\{f_1,\de_{e_a}.f_2\}_M\ +\ f_a^{bd} (\de_{e_b}.f_1)(\de_{e_d}.f_2)\ =\
\de_{e_a}.\{f_1,f_2\}_M \label{EIxi}
\end{eqnarray}
It follows immediately from eq.(\ref{EIxi}) that a Lie-Poisson action cannot
be Hamiltonian unless the algebra $\CH^*$ is Abelian.
 However, in general, we have a non-Abelian analogue of
the Hamiltonian action eq~.~(\ref{EIiii})~\cite{Lu90}:
There exists a function
$\Ga$, locally defined on $M$ and taking values in the group $H^*$,
such that for any function $f$ on $M$,
\begin{eqnarray}
\de_X.f\ =\ <\ \Ga^{-1}\ \{f,\Ga\}_M,X>\quad,\quad \forall\ X\in\CH
\label{EIxii}
\end{eqnarray}
We refer to $\Ga$ as the non-Abelian Hamiltonian of the Lie-Poisson
action.
The moment map $\CP$ for the Lie-Poisson action is the map
$\CP$ from $M$ to $H^*$ defined by $x\longrightarrow \Ga(x)$.\par
The proof is the following. Introduce the Darboux coordinates $(q^i,p^i)$.
Let $\Om= e^a\ \Om_a$ be the $\CH^*$-valued one-form
defined by $\Om_a=e_a^{q^i}dp^i\ -\ e_a^{p^i}dq^i$ where
$e_a^{q^i}$, $e^{p^i}_a$ are the components of the vector field $e_a$,
$e_a=e_a^{q^i}\d_{q^i}+e_a^{p^i}\d_{p^i}$. Eq. (\ref{EIxi}) is then
equivalent to the following zero-curvature condition for $\Om$:
$d\Om\ + \[\Om,\Om\]_{\CH^*}\ =\ 0 $.
Therefore, locally on $M$, $\Om=\Ga^{-1}\ d\Ga$.
This proves eq.(\ref{EIxii}).
The converse is true: an action generated by a non-Abelian
Hamiltonian as in eq.(\ref{EIxii}) is Lie-Poisson since then we have:
\begin{eqnarray}
\de_X.\{ f_1 ,f_2 \}_M &-& \{\de_X.f_1 , f_2 \}_M
- \{f_1 , \de_X.f_2 \}_M \nonumber \\
&=& <\BBL \Gamma^{-1} \{ f_1 ,\Ga \}_M ~,~
\Ga^{-1}\{f_2 ,\Ga \}_M \BBR _{\CH^*} ,X > \nonumber
\end{eqnarray}

The non-Abelian symmetries we descrided above thus provide an exemple
of Lie-Poisson actions, since they are generated by a non-Abelian
Hamiltonian. They actually are a particular exemple
of more general transformations called dressing transformations
in solitons theories \cite{ZaSh79,Se85}.

\section{The quantum Heisenberg chain}

We introduce the quantum Heisenberg chain as a quantization
of the discrete analogue of the  classical Heisenberg model.
Since, it is not more difficult to deal with the $su(p)$ algebra,
we generalize it from $su(2)$ to $su(p)$.
So let us assume that we have discretized the interval on which the
model was defined into $N$ segments of identical length that we denote
by $h$. The extremities of the segments form a chain of sites on which
the $su(p)$ spin variables leave. We denote the $su(p)$ spin variables at
the sites $j$ by $S_j^{ab}$ with $a,\ b =1,\cdots,p$. They satisfy the
commutation relations of the discretized $su(p)$ loop algebra:
\debut
\BBL\ S^{ab}_j\ ,\ S^{cd}_k\ \BBR = \de_{jk}
\BL\ \de^{cb}\ S^{ad}_j - \de^{ad}\ S^{cb}_j\ \BR \label{CIIi}
\fin
This corresponds to the quantization of the Poisson brackets (\ref{Ki}).
Approximating the derivatives by finite differences,
and using the constraints $S^2_k=s^2={\rm const}$,
the discretized version of the Hamiltonian (\ref{Kii}) becomes,
(up to a constant term):
\debut
H = \sum_{k=1}^{N}\sum_{ab} S^{ab}_k S^{ba}_{k+1}
\label{CIIii}
\fin
Here, we have assumed periodic boundary conditions.
As is well known, in order to preserve the integrability we have to
choose the  spin operators $S^{ab}_k$ to act on the fundamental vector
representation of $su(p)$. So on each sites there is a copy of
$\C^p$ and the operator $S^{ab}_j$  which acts only the $j^{th}$ copies of
$\C^p$ is represented by the canonical matrice $\ket{a}\bra{b}$.

In the discretized quantized model, the monodromy matrix, which is a
$p \times p$ matrix with operator entries, has matrix elements:
\debut
T_{ab}(\lambda)&=&P \exp\left[-\int A_x(y,\lambda) dy
 \right]_{ab}^{ {\rm discretized} } \non
 &=&  \sum_{a_1\cdots a_{N-1}}\
\BBL{1 + \frac{h}{\lambda} S_1}\BBR^{a a_1}
\BBL{1 + \frac{h}{\lambda} S_2}\BBR^{a_1 a_2}\cdots
\BBL{1 + \frac{h}{\lambda} S_N}\BBR^{a_{N-1} b}
\label{CIIIiii}
\fin
As for the classical theory, the
transfer matrix admits an $\inv{\lambda}$ expansion:
\debut
T_{ab}(\lambda) = \delta_{ab} + \frac{h}{\lambda}\sum_k S^{ab}_k
 +\frac{h^2}{\lambda^2} \sum_{j<k} \sum_d S^{ad}_j S^{db}_k +\cdots
\nonumber
\fin

The important properties of this monodromy matrix is that we
can compute the commutation relations between its matrix elements.
These relations can be gathered into the famous relations of the
algebraic Bethe anstaz, see e.g. \cite{Fa82,Ga0}:
\debut
R(\lambda-\mu) (T(\lambda)\otimes 1) (1\otimes T(\mu))
=( 1\otimes T(\mu))(T(\lambda)\otimes 1) R(\lambda-\mu)
\label{CIIiv}
\fin
where $R(\lambda)$ is Yang's solution of the Yang-Baxter equation:
\debut
R(\lambda) = \lambda - h\ P \nonumber\fin
with $P$ the exchange operator $P(x\otimes y)= y \otimes x$.
For the matrix elements of the monodromy matrix this becomes:
\debut
(\lambda-\mu) \BBL T_{ab}(\lambda), T_{cd}(\mu)\BBR =
h \BL T_{ad}(\lambda)T_{cb}(\mu) - T_{ad}(\mu) T_{cb}(\lambda) \BR
\label{CIIva}
\fin
These relations are the quantum analogues of the Poisson brackets (\ref{EHix}).
As in the classical theory, it implies that the trace of the
monodomy matrix, or more accurately the logarithm
of the trace of the monodromy matrix,
 form a generating function of commuting quantities.
In the same as in the previous section, the
local Hamiltonians are not found by expanding around $\lambda=\infty$
but around $\lambda=0$. In particular, we recover the Hamiltonian
by expanding the logarithm of the trace to first order:
$ H \propto \partial_\lambda \log \lambda^NT(\lambda)
\Big\vert_{\lambda=0}$.

\subsection{Non-Abelian symmetries and the quantum Yangians.}

The Hamiltonian is clearly $su(p)$ invariant. But, in the same as
for the classical theories, the symmetry group is much bigger.
The naive discrete quantum analogue of the classical charges $Q^0$
and $Q^1$ introduced  in eq. (\ref{CIv}) is:
\debut
Q^0_{ab} &=& \sum_k S^{ab}_k  \non
Q^1_{ab} &=& \frac{h}{2} \sum_{j<k} \sum_d
(S^{ad}_jS^{db}_k - S^{ad}_k S^{db}_j)
\label{CIIvi}
\fin
This naive ansatz turns out to be correct:
the charges $Q^0_{ab}$ commutes with the Hamiltonian and the
charges $Q^1_{ab}$ formally commutes for chains of infinite
length. For finite chain the commutation is broken by boundary
terms. However, the corresponding currents, which are conserved on
the lattice (of any size), were constructed in \cite{BF91}.

The charges (\ref{CIIvi}) form a non-Abelian algebra, which is
not a Lie algebra.
They satisfy the following commutations relations:
\debut
\BBL Q^0_{ab},Q^0_{cd} \BBR &=& \de_{cb} Q^0_{ad} - \de_{ad} Q^0_{cb}\non
\BBL Q^0_{ab},Q^1_{cd} \BBR &=& \de_{cb} Q^1_{ad} - \de_{ad} Q^1_{cb}
\label{CIIix}\\
\BBL Q^1_{ab},Q^1_{cd} \BBR &=& \de_{cb} Q^2_{ad} - \de_{ad} Q^2_{cb}
 + \frac{h^2}{4} Q^0_{ad}(\sum_e Q^0_{ce}Q^0_{eb}) -
 \frac{h^2}{4} (\sum_e Q^0_{ae}Q^0_{ed})Q^0_{cb} \nonumber
\fin
Here $Q^2_{ab}$ is a new operator whose explicit expression is
irrelevent for our discussion.
The remarkable fact is that the extra non-linear term in the last
equation can be expressed only in terms of $Q^0_{ab}$. This last
equation implies a relation involving only $Q^0_{ab}$
and $Q^1_{ab}$:
\debut
\BBL Q^0_{ab},\BBL Q^1_{cd},Q^1_{ef}\BBR\BBR &-&
\BBL Q^1_{ab},\BBL Q^0_{cd},Q^1_{ef}\BBR\BBR \non
= \frac{h^2}{4} \sum_{p\, q}\BL
\BBL Q^0_{ab},\BBL Q^0_{cp}Q^0_{pd},Q^0_{eq}Q^0_{qf}\BBR\BBR &-&
\BBL Q^0_{ap}Q^0_{pb},\BBL Q^0_{cd},Q^0_{eq}Q^0_{qf}\BBR\BBR \BR
\label{CIIIx}
\fin
The associative algebra generated by elements $Q^0_{ab}$ and
$Q^1_{ab}$ satisfying the relations (\ref{CIIix}) and (\ref{CIIIx})
is called the $su(p)$ Yangians \cite{Dr86}. As it can be seen by comparing
with eq. (\ref{CIx}), this algebra is a deformation of the $su(p)$
loop algebra.

The $su(p)$ Yangian is not an Lie algebra but a Hopf algebra.
It is therefore equipped with a comultiplication $\De$, which
is a homomorphism from the algebra into the tensor product
of two copies of the same algebra. For the $su(p)$ Yangians,
the comultiplication is given by:
\debut
\De Q^0_{ab} &=& Q^0_{ab}\otimes 1 + 1 \otimes Q^0_{ab} \label{Ecom}\\
\De Q^1_{ab} &=& Q^1_{ab}\otimes 1 + 1 \otimes Q^1_{ab}
 +\frac{h}{2} \sum_d\BL Q^0_{ad}\otimes Q^0_{db} -
 Q^0_{db}\otimes Q^0_{ad} \BR \nonumber
\fin
It can be used to construct tensor products of representations.
For $h=0$ it reduces to the comultiplication of the $su(p)$
loop algebra.

\subsection{The quantum transfer matrix and an alternative presentation
of the Yangians.}

The charges $Q_{ab}^0$ and $Q^1_{ab}$ appears as the first terms in the
$\inv{\lambda}$ expansion of the quantum monodromy matrix $T(\lambda)$.
Therefore, the $su(p)$ Yangian can also be presented in terms
of $T(\lambda)$, or more precisely, in terms of its components
in an $\inv{\lambda}$ expansion:
\debut
T_{ab}(\lambda) = \de_{ab} + h \sum_{n=0}^\infty \lambda^{-n-1} t^{(n)}_{ab}
\label{CIIxi}
\fin
Therefore, the alternative presentation of
the $su(p)$ Yangian is as the associative algebra generated by the
elements $t^{(n)}_{ab}$ with relations:
\debut
\BBL  t^{(n)}_{ab},t^{(m)}_{cd}  \BBR
= \delta_{cb}t^{(n+m)}_{ad}- \delta_{ad} t^{(n+m)}_{cb}
+ h\sum_{p=0}^{n-1}\BL  t^{(m+p)}_{ad}t^{(n-1-p)}_{cb}
- t^{(n-1-p)}_{ad}t^{(m+p)}_{cb} \BR \label{CIIxii}
\fin
These relations are equivalent to :
\debut
\BBL  t^{(0)}_{ab},t^{(m)}_{cd}  \BBR
&= &\delta_{cb}t^{(m)}_{ad}- \delta_{ad} t^{(m)}_{cb} \label{CIIxiii}\\
\BBL  t^{(n+1)}_{ab},t^{(m)}_{cd}  \BBR
 -\BBL  t^{(n)}_{ab},t^{(m+1)}_{cd}  \BBR
&=& h\BL  t^{(m)}_{ad}t^{(n)}_{cb}
- t^{(n)}_{ad}t^{(m)}_{cb} \BR \nonumber
\fin
which, in their turn, are equivalent to the fundamental commutation
relations (\ref{CIIiv}).
In the quantum theory, the unit determinant constraint is modified
into the condition that the so-called quantum determinant of the
transfer matrix is one:
\debut
Det_q\ T(\lambda) \equiv \sum_{\sig {\rm perm.}} \ep(\sigma)
T_{\sigma(p) p}(\lambda) ..... T_{\sigma(1) 1}(\lambda+ph) = 1
\label{CIIxiv}
\fin
The sum is over the permutation of $p$ objects and $\ep(\sig)$ is
the signature of the permutation $\sigma$.
The quantum determinant commutes with all the components of the
monodromy matrix, so this constraint can be imposed in a consistent
way.

With the quantum determinant constraint,
the $\inv{\lambda}$ expansion of the
monodromy matrix can be reconstructed from its
two first components $t^{(0)}_{ab}$ and $t^{(1)}_{ab}$.
Finally, the relation between these two components and
the quantum charges $Q^0_{ab}$ and $Q^1_{ab}$ is:
\debut
Q^0_{ab} &=& t^{(0)}_{ab} \non
Q^1_{ab} &=& t^{(1)}_{ab} - \frac{h}{2} \sum_d t^{(0)}_{ad} t^{(0)}_{db}
\label{CIIxv}
\fin
This shows that, as for the classical theory, the knowledge of the
first two non-local charges $Q^0_{ab}$ and $Q^1_{ab}$ is equivalent ot
the knowledge of the $\inv{\lambda}$ expansion of the monodromy matrix.

The comultiplication for the transfer matrix is given by:
\debut
\De T_{ab}(\lambda) = \sum_d T_{ad}(\lambda)\otimes T_{db}(\lambda)
\nonumber
\fin
For $Q^0_{ab}$ and $Q^1_{ab}$ it reduces to eq. (\ref{Ecom}).
The adjoint action of the transfer matrix on an operator $\Phi$ is:
\debut
\BBL {\rm Adj.} T_{ab}(\lambda)\BBR \Phi = \sum_d T_{ad}(\lambda) \
\Phi \ T_{db}^{-1}(\lambda) \nonumber
\fin
where $T_{ab}^{-1}(\lambda)$ is the matrix of operators characterized
by $\sum_dT_{ad}(\la) T^{-1}_{db}(\la)=\de_{ab}$. This adjoint action
is the quantum analogue of the semi-classical equation (\ref{CIxvi}).

\subsection{Digression on the double Yangians.}

The Yangians are deformation of only half of the loop algebras,
i.e. they are deformation of only the sub-algebra of loops
regular at the origin. A quantum deformation of the complete
loop algebra can be obtained by introducing the quantum
double of the Yangian, following Drinfel'd \cite{Dr86}.
In quantum field theory, the double Yangian was introduced
in \cite{LS91}. It can be presented in the following multiplicative
form similar to eq. (\ref{CIIiv}). Let $T^+_{ab}(\la)$
and $T^-_{ab}(\la)$ be two $p\times p$ matrices of operators
with the following expansion:
\debut
T^+_{ab}(\lambda) &=& \de_{ab} +
 h \sum_{n=0}^{+\infty} \lambda^{-n-1} t^{(n)}_{ab}\non
T^-_{ab}(\lambda) &=& h \sum_{n=0}^{+\infty} \lambda^{n} t^{(-n-1)}_{ab}
\nonumber
\fin
Note that these expansions correspond to expansions into
series regular at the origin or at infinity.
The $gl(p)$ double Yangian is the algebra generated by the
elements $t^{(\pm n)}_{ab}$ with relations:
\debut
R(\lambda-\mu) (T^\pm(\lambda)\otimes 1) (1\otimes T^\pm(\mu))
&=&( 1\otimes T^\pm(\mu))(T^\pm(\lambda)\otimes 1) R(\lambda-\mu)\non
R(\lambda-\mu) (T^+(\lambda)\otimes 1) (1\otimes T^-(\mu))
&=&( 1\otimes T^-(\mu))(T^+(\lambda)\otimes 1) R(\lambda-\mu)
\nonumber
\fin
If in a finite dimensional representation of the Yangian the transfer matrix
$T^+_{ab}(\lambda)$ is a polynomial of degree $N$ in $1/\lambda$, then
the double Yangian is represented in the same vector space by
$T^-_{ab}(\la)\propto \la^N T^+_{ab}(\la)$.

\end{document}